\documentclass[aps,pre,twocolumn,amsmath,amssymb,floatfix]{revtex4-1} 

\usepackage{graphicx}
\usepackage{amsmath}
\usepackage{comment}
\usepackage{color}
\usepackage{bm}

\begin{document}

\title{Unusual liquid phases for indented colloids with depletion interactions}

\author{Douglas J. Ashton}
\author{Robert L. Jack}
\author{Nigel B. Wilding}
\affiliation{Department of Physics, University of Bath, Bath BA2 7AY, United Kingdom}
\date{\today}

\newcommand{\epsLK}{\varepsilon_{\rm LK}}
\newcommand{\epsBB}{\varepsilon_{\rm BB}}

\newcommand{\etas}{\eta_{\rm s}}

\newcommand{\rhoL}{\rho_{\rm L}}
\newcommand{\rhoK}{\rho_{\rm K}}
\newcommand{\rhoLK}{\rho_{\rm LK}}
\newcommand{\rhoLLK}{\rho_{\rm LLK}}
\newcommand{\rhoLKK}{\rho_{\rm LKK}}
\newcommand{\rhoKK}{\rho_{\rm KK}}

\newcommand{\rhotree}{\rho_{\rm tree}}

\newcommand{\rlj}[1]{{\color{blue}#1}}
\newcommand{\nigel}[1]{{\color{green}}}



\begin{abstract}

We study indented spherical colloids, interacting via depletion forces.
These systems exhibit liquid-vapor phase transitions whose properties
are determined by a combination of strong ``lock-and-key'' bonds
and weaker non-specific interactions.  
As the propensity for lock-and-key
binding increases, the critical point moves to significantly lower density, 
and the coexisting phases change their structure.
In particular, the liquid phase is ``porous'', exhibiting large percolating voids.
The properties of this system depend strongly on the topological structure of 
an underlying bond network: {we comment on the implications of this fact
for the assembly of equilibrium states with controlled porous structures}.

\end{abstract}

\maketitle

In colloidal systems, a wide range of structures can be self-assembled
from a simple palette of components and
interactions~\cite{Glotzer2007}.  For example, anisotropic attractive
forces between particles can stabilise colloidal micelles or exotic
crystals
~\cite{Wang:2012uq,Schade:2013fk,Sacanna:2010ys,Rossi:2011qd,Chen:2011eu,Sacanna:2012qe},
which could form the basis of future materials and devices.
Anisotropic attractions can be achieved experimentally via chemical
patches~\cite{Kraft:2009jt,Kraft:2011cs,Chen:2011eu,Leunissen:2009lr,Wang:2012uq},
or through a combination of particle shape and depletion
forces~\cite{Xia:2012zg,Gantapara:2013to,Sacanna:2013kq,Sacanna:2010ys,Rossi:2011qd}.
Depletion forces arise when colloids are dispersed in a solution of
much smaller ``depletant'' particles, which induce an attractive
interaction~\cite{Lekkerkerker:2011}, whose properties can be tuned
via the size and number density of the depletant.  Here, we present
Monte Carlo simulations of the assembly of indented
(``lock-and-key'') colloidal
particles~\cite{Sacanna:2010ys,Sacanna2011lock,odriozola:111101,Ashton2013,Ashton2014oxford},
in the presence of an ideal depletant~\cite{Asakura1954}.  We introduce an effective
potential which captures quantitatively the effect of this depletant.
This enables efficient simulation, and accurate characterisation of
the liquid-vapor phase transitions that occur in this system. When
the depletant particles are small in size compared to the colloids,
the resulting liquid and vapor phases have unusual properties.  In
particular, the critical point for the phase transition occurs at a
rather low density, and the liquid phases have a complex structure
that includes large voids.  Indeed, we find that the liquid phase is
``porous'' on a length scale comparable with the collidal particles.
Similar properties have also been found in model polymer
liquids~\cite{Wilding1996c,Frauenkron:1997ng,pana1998} and in the ``empty
liquids'' that occur in patchy-particle
models~\cite{Bianchi2006,Bianchi2008,Ruzicka:2011ud}.  However, the
situation for indented colloids is different from both these cases: we
trace these differences to the topology of the bond networks that
appear as self-assembly takes place.

\begin{figure}
\begin{center}
\includegraphics[width= 0.9 \columnwidth]{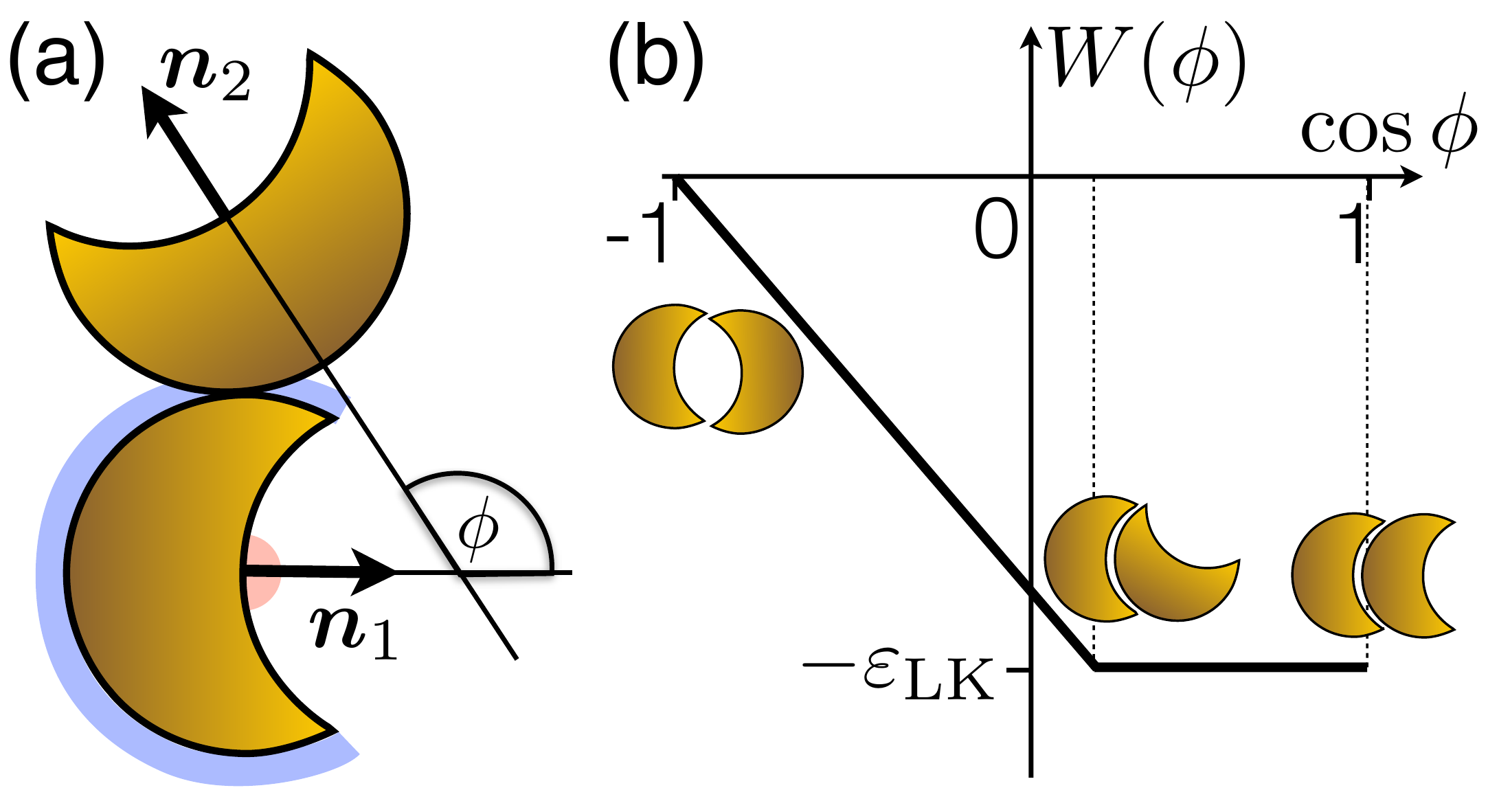}
\caption{
Illustration of the effective potential between indented colloids.
{(a)}~The interaction depends on the vector $\bm{r}_{12}$ between the
colloids, and their orientations $\bm{n}_{1,2}$.  We define two angles
$\theta_{\rm R}$ and $\phi$ by $\cos\theta_{\rm R} = \max(
\bm{n}_1\cdot \bm{r}_{12},-\bm{n}_2\cdot \bm{r}_{12})$ and $\cos\phi =
\bm{n}_1\cdot \bm{n}_2$.  Particles can bind in a lock-key
configuration (small $\theta_{\rm R}$) or a back-to-back configuration
(large $\theta_{\rm R}$), as indicated by the two shaded areas.
(b)~In the lock-key orientation, the interaction depends on the angle
$\phi$ through the function $W(\phi)$, the form of which is motivated
by studies of the exact depletion potential (see Appendix A). 
}
\label{fig:model}
\end{center}
\end{figure}

\begin{figure*}
\includegraphics[width=17.5cm]{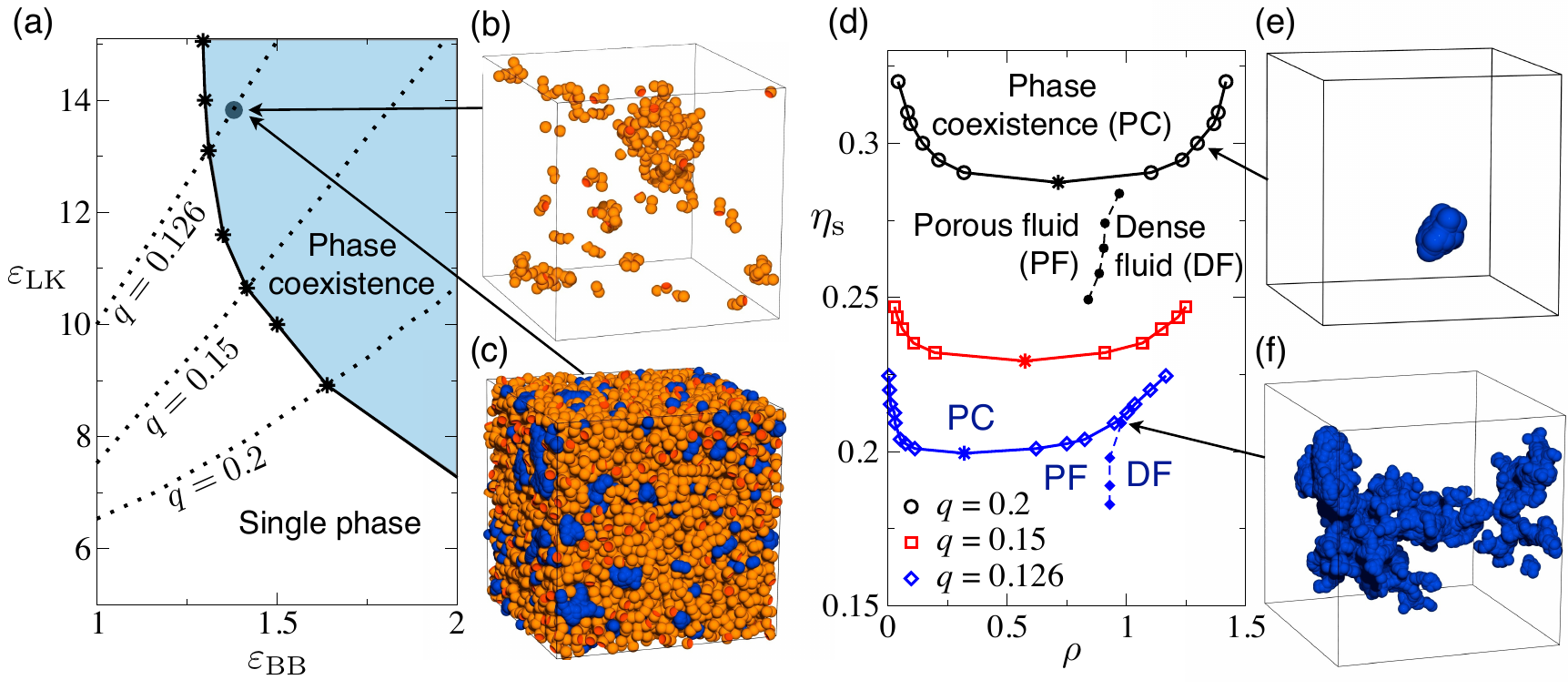}
\caption{(a)~Summary of phase behaviour, as a function of $\epsLK,\epsBB$.  In the shaded region, coexistence of liquid
and vapor phases can occur, for suitable colloid densities $\rho$.  The stars indicate the positions of critical points. 
 The dotted lines show how the effective interaction strengths $(\epsLK,\epsBB)$ depend on the depletant size ratio $q$ 
 (the depletant volume fraction $\eta_{\rm s}$ varies along these lines).  (b,c) Representative configurations from
the  coexisting
``liquid'' and ``vapor'' phases at the indicated state point, which corresponds to $\eta_{\rm s}=1.05\eta_{\rm s}^c$. 
 For the liquid state, the void space is colored blue, to emphasise the
porous structure.  (d)~Binodal curves showing the densities of coexisting liquid and vapor phases as a function of depletant parameters
$q,\etas$ and (total) colloid density $\rho$. Stars mark critical points. The dashed lines show
the densities at which the void spaces percolate the system, as discussed in the main text.  (e,f) Visualisations of the largest void spaces
from representative configurations at the indicated state points, both of which correspond to $\eta_{\rm s}=1.05\eta_{\rm s}^c$.%
}
\label{fig:liq-vap}
\end{figure*}

An indented colloidal particle is modelled as a
sphere of diameter $\sigma$, from which we
cut away its intersection with a
second sphere of the same diameter~\cite{Ashton2013,Ashton2014oxford}. The distance between the sphere
centres is taken as $d_c=0.6\sigma$, so the depth of the indentation (measured from the ``lip'') 
is $0.2\sigma$, comparable with experimentally realizable
particles \cite{Sacanna:2010ys}. The orientation of particle $i$ is specified by a unit vector $\bm{n}_i$
that points outwards through the centre of the indentation.
The hard particle interactions between these particles are treated exactly in our model, but
we parameterise the depletion interaction between the colloids through an effective potential,
illustrated in Fig.~\ref{fig:model}.
Specifically, when two colloids approach one another in the
``back-to-back'' configuration, they feel a square-well interaction
potential of depth $\epsBB$ and range $r_{\rm BB}$. When they approach
in the ``lock-key'' configuration, the interaction range is $r_{\rm
  LK}$, and the effective potential also depends on the angle $\phi$
between the particle directors.  For perfect lock-key binding, the
depth of the potential well is $\epsLK$.  We fix the unit of energy by
setting $\beta=1/(k_{\rm B}T)=1$.  The interaction ranges are fixed
throughout this work according to $(r_{\rm LK}-d_c)=(r_{\rm
  BB}-\sigma)=0.1\sigma$, comparable with the size of the depletant
particles.  (As for spherical particles~\cite{Noro2000}, the behavior
depends weakly on these interaction ranges, if the interaction
strengths are adjusted so that the binding free energies or second
virial coefficients for pair interactions are held constant, and the
ranges are small compared to $\sigma$.)  To obtain the interaction
strengths $\epsLK$ and $\epsBB$ as a function of depletant parameters,
we used the geometric cluster algorithm (GCA)~\cite{Dress1995,Liu2004} to
simulate systems of two colloidal particles interacting with a
depletant of penetrable spheres~\cite{Asakura1954}.  The size ratio between colloids and
depletant particles is $q$ (the depletant diameter is $q\sigma$), and
the depletant volume fraction is $\etas$.  Full details of the
effective potential and the parameterisation of ($\epsLK,\epsBB)$ are
given in Appendix A. 

We employed grand canonical Monte Carlo (GCMC) simulation to study the
phase behaviour~\cite{Wilding1995,Bruce1992} of the colloids,
interacting through the effective potential, in cubic boxes of sizes between $(12\sigma)^3$ and
$(20\sigma)^3$. Smaller boxes sufficed for large $q$,
where the size of bound clusters is typically small; bigger
boxes were needed to accommodate the larger clusters that form at small
$q$.  Our GCMC method uses the usual particle updates (insertions,
deletions, displacements and rotations), combined with ``biased''
insertions and deletions, which attempt to add (or remove) a colloid
in a lock-key bound state, subject to satisfying detailed balance. This
innovation, (described further in Appendix B), 
increases the efficiency of
the simulation by up to 4 orders of magnitude.

Fig.~\ref{fig:liq-vap}(a) summarises the phase behavior of the indented colloids.
The state point of the system is specified by three parameters
$(\rho,\epsLK,\epsBB)$ where $\rho$ is the number density of colloids.  
The solid line in Fig.~\ref{fig:liq-vap}(a) indicates
values of $(\epsLK,\epsBB)$ for which a liquid-vapour critical point
exists, at some critical density $\rho^{\rm c}$. For values of
$(\epsLK,\epsBB)$ above this line, there exist values of the density for which
liquid-vapor phase coexistence occurs: see for example
Figs.~\ref{fig:liq-vap}(b,c).  The structure of these phases is
discussed further below.  Dotted lines in Fig.~\ref{fig:liq-vap}(a)
show how the parameters of the effective model are related to
depletant parameters ($q,\etas$): for smaller $q$, lock-key binding is
more favorable than back-to-back binding.  Increasing $\etas$ at
constant $q$ leads to an increase in both $\epsLK$ and $\epsBB$, 
along the dotted lines.  To assess the accuracy of the
effective potential, we used the GCA within a restricted Gibbs ensemble
\cite{Ashton:2010ys} to locate the critical point in a system of indented colloids with
explicit depletant, for $q=0.2$.  {We find that the critical parameters 
lie within $10\%$ of the results obtained with the effective potential.}
However, with explicit depletant, it was not possible to investigate the
structure of liquid and vapor phases, nor the behavior for smaller
$q$, due to the computational cost required. Hence our use of the
effective potential in this work.

The binodal curves associated with liquid-vapor phase coexistence are shown in Fig.~\ref{fig:liq-vap}(d), as a function
of the depletant volume fraction $\etas$, for three values of $q$.  As the depletant particles get smaller (decreasing $q$ at fixed $\etas$),
the attractive forces between colloids get stronger, so phase separation occurs for
smaller values of $\etas$.  However, the most striking effect in Fig.~\ref{fig:liq-vap}(d) is the strong decrease (by more than a factor of $2$)
in the critical density of the colloids $\rho^c$.  The origin of this effect is the very strong lock-key binding that occurs when $q$ is small.

Figs. \ref{fig:liq-vap}(b,c) illustrate the unusual structures of the coexisting phases for $q=0.126$, corresponding to strong lock-key binding.
There are large void spaces within the liquid: we identify these by inserting spherical
``ghost'' particles of size $\sigma$ into the system, wherever this is possible without
overlapping an existing colloid (ghost particles may overlap with each other).  
These particles, shown in Fig.~\ref{fig:liq-vap}(c), highlight the existence of
the voids.  At this state point, the depletant volume fraction is $5\%$ greater than its critical value $\eta_s^c$, 
so the system is significantly outside
the critical regime: these voids are not associated with critical fluctuations, but are intrinsic to the 
liquid states that occur for small $q$.  

To reinforce the unusual nature of the liquid state at small $q$, we note that  liquid configurations for larger $q$ lack these large voids: they
resemble ``normal'' colloidal liquids of spheres. The largest void in a typical liquid configuration at $q=0.2$
and $\eta_s = 1.05\eta_s^c$ is shown in Fig.~\ref{fig:liq-vap}(e), and is much smaller than the corresponding void in Fig.~\ref{fig:liq-vap}(f).  The ghost particles described above allow
this effect to be analysed quantitatively: the typical volumes associated with the voids are shown in Appendix C. 
We have also analysed the state points at which the ghost particles
percolate the system, as shown by the ``void percolation'' lines in Fig.~\ref{fig:liq-vap}(d).  On the low-density side of
these lines, the voids form a percolating network, and we expect that a test particle of a size comparable with a colloid can travel freely 
through the liquid: this may be a useful criterion for identifying ``empty''~\cite{Bianchi2006,Bianchi2008,Ruzicka:2011ud} or ``porous'' liquids.  At higher densities,
the liquid resembles more familiar colloidal liquids.  The key point to be noted from Fig.~\ref{fig:liq-vap}(d) is that decreasing $q$ leads
to an increasing range of density ($\rho$) and interaction strength ($\etas$) for which porous liquids exist.

The unusual structure of these liquid and vapor states originates from a
hierarchy of energy scales -- the system supports strong lock-key
bonds, and weaker back-to-back binding.  To illustrate this effect,
imagine that the colloids form linear chains, connected by strong
lock-key bonds~\cite{Ashton2013}.  For large $\epsLK$, almost all of
the available lock-sites are participating in binding -- the weaker
back-to-back binding then provides both inter-chain and intra-chain
interactions, which can cause the chains to aggregate or collapse, as
happens in solutions of
polymers~\cite{Grosberg,Wilding1996c,Frauenkron:1997ng,pana1998}.  However,
chains of colloids linked by depletion forces are in fact akin to
``living polymers''~\cite{Zheng1992},
in that they are continuously breaking and
re-forming in the equilibrium state.  One can associate the liquid
state with an aggregated state of many polymers, while the unusual
vapor phase contains polymeric chains that tend to collapse into
compact states [recall Fig.~\ref{fig:liq-vap}(b)].

In fact, Fig.~\ref{fig:liq-vap}(d) closely resembles
the behaviour of polymers in solution, interacting via a relatively weak non-specific attraction~\cite{Wilding1996c,Frauenkron:1997ng,pana1998}. 
As the polymer length increases in that system,
the critical point moves to lower polymer density and weaker non-specific interactions.  
These systems also exhibit strong ``field-mixing'' at the critical point~\cite{Bruce1992,Wilding1995}: the natural order
parameter for the liquid-vapor phase transition is not the density of the system, but rather
a linear combination of the density and the energy.  We show evidence in Appendix B 
that the liquid-vapor critical points
of the indented colloids are in the Ising universality class (as expected), but with strong field-mixing, as in the polymer case ~\cite{Frauenkron:1997ng}.
{This latter effect reflects the unusual structures of the liquid and vapor states, especially the strong asymmetry between the structures of liquid and vapour states in
Fig.~\ref{fig:liq-vap}(b,c), (that is, particle-hole asymmetry).}

This analogy with polymers leads to a prediction for the colloidal system:
in the limit of long polymers, one expects the liquid-vapor critical point to occur at the $\Theta$-temperature of the 
polymer~\cite{Wilding1996c,Frauenkron:1997ng,pana1998}, where the chain statistics are those of a simple random walk.
This requires a weak non-specific attraction, to overcome the excluded volume interactions that cause the polymer 
to swell. For the colloidal system, the long-polymer limit corresponds to $\epsLK\to\infty$, in which case one would
expect the liquid-vapor critical point to occur when the non-specific (back-to-back) interaction $\epsBB$ balances the excluded
volume swelling effect.  This implies that the line of critical points in Fig.~\ref{fig:liq-vap}(a) should
tend to a non-zero value of $\epsBB$, as $\epsLK\to\infty$, consistent with our results.

\begin{figure}
\begin{center}
\includegraphics[width= 0.98 \columnwidth]{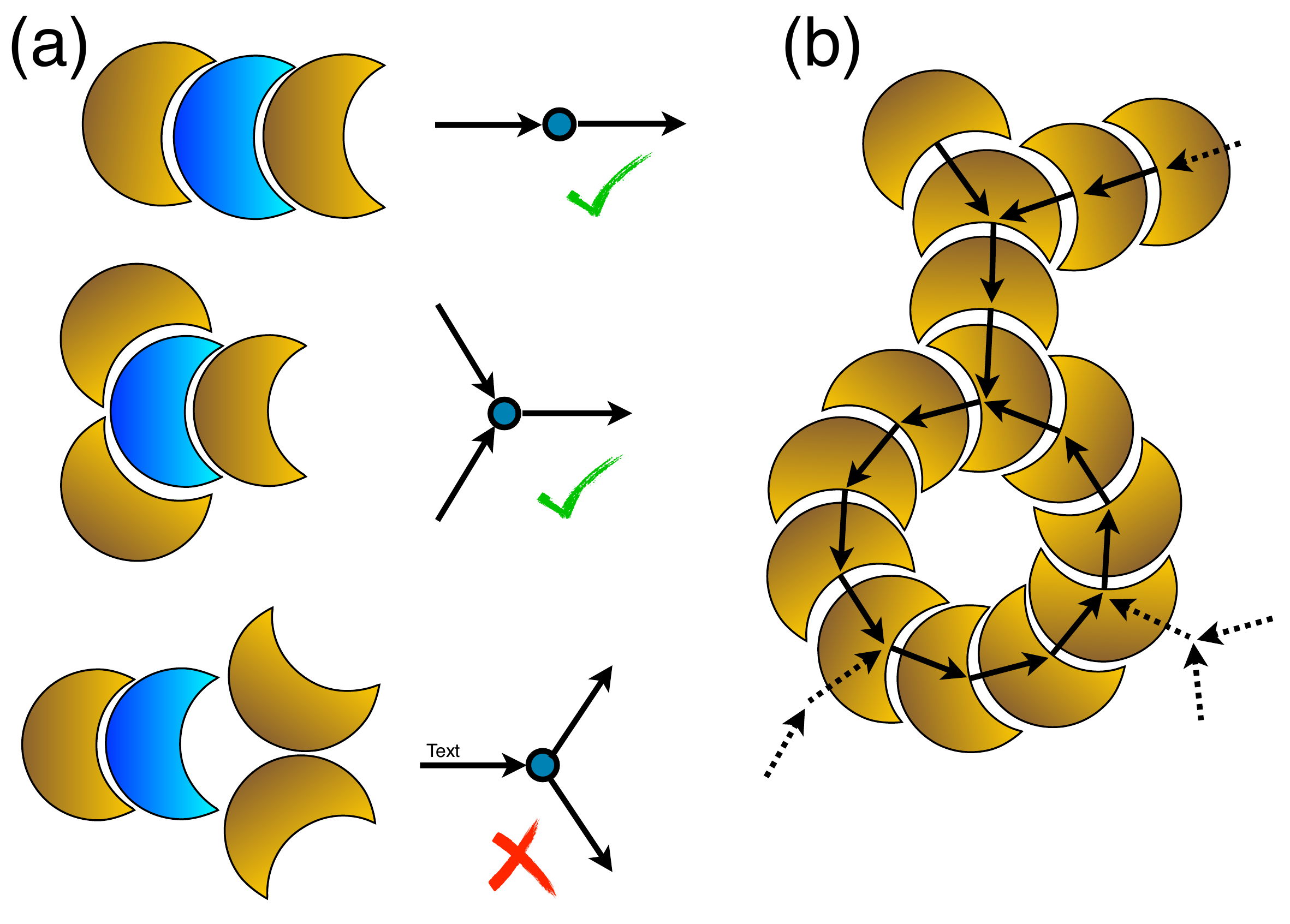}
\caption{(a)~Network motifs for lock-key bonds. Circles
indicate particles and arrows indicate lock-key bonds, directed from the ``lock'' to the ``key''.  The most
important local environments are linear chains (one outward and one inward bond) and branching points
(two inward and one outward bond).  The converse branching case (one inward and two outward bonds)
is not possible, due to the particle shape.
(b)~Clusters of particles can be characterised by networks of arrows.
Once a connected cluster of particles includes a single loop, it can grow further only by forming inward bonds
(some possible locations for additional bonds are shown with dotted arrows).
It follows that each connected cluster can include at most one loop.}
\label{fig:network}
\end{center}
\end{figure}

However, while this analogy between chains of indented colloids and linear living polymers is appealing, it misses
an important feature of the colloidal system.  For the particle shape considered here,
these colloids readily form \emph{branched} chains.  In patchy-particle models, the presence 
of branching is sufficient to drive a liquid-vapour phase transition~\cite{Bianchi2006,Bianchi2008,Ruzicka:2011ud}, but this is not the
case for lock-and-key colloids.  The reason is the directed nature of the lock-key bond.
In Fig.~\ref{fig:network}, we illustrate different local
structures in these systems, and the construction of a bond network.  
Lock-key bonds are indicated by arrows that point from the ``lock'' particle to the ``key''.
In this representation, the crucial feature is that a particle may have several inward bonds, but 
at most one outward bond. It follows 
that a cluster of particles connected by lock-key bonds may
contain at most one loop, as illustrated in Fig.~\ref{fig:network}(b).

This feature has important implications for phase behaviour -- a condensed liquid phase can be stable
only in the presence of percolating clusters that contain large numbers of internal loops.  This effect can be illustrated using
Wertheim's theory of associating
fluids~\cite{Wertheim3,Bianchi2006,Bianchi2008,Ashton2013}, which divides particles into ``quasispecies'' according
to their local bonding, and predicts the number densities associated with these species.  For indented colloids, 
we define $\rhoL$ as the
number density of unoccupied ``lock sites'' (colloidal indentations), and $\rhoK$ as the number density
of free ``key sites'' (concave colloid surfaces) to which no particle is bound.  The theory provides ``mass-action''
equations which relate the numbers of bonds in the system to $\rhoL$ and $\rhoK$~\cite{Ashton2013}: see Appendix D. 

Within Wertheim's theory, the pressure $P$ of a fluid of clusters satifies~\cite{Bianchi2008}
\begin{equation}
\beta P = \rho_{\rm tree} + \Delta P_{\rm ref}
\end{equation}
where $\rhotree$ is an estimate of the
number density of bonded clusters
in the system, and $\Delta P_{\rm ref}$ is a contribution coming from a hard-particle 
reference system.  
The density $\rho_{\rm tree}$ counts the number of clusters under the assumption that they
are all ``tree-like'' (containing no loops).  For indented colloids without back-to-back interactions,
we have $\rhotree = \rhoL$ (in the absence of loops, every cluster has exactly one unoccupied lock site).
However, liquid-vapor phase transitions require $(\partial P/\partial \rho)<0$; 
it is easily shown within Wertheim's theory 
that $(\partial \rhoL/\partial \rho)>0$, and we expect on
very general grounds that $\Delta P_{\rm ref}$ is increasing in $\rho$.  Hence, in the absence of back-to-back binding,
the theory predicts that no phase transition is possible: see Appendix D.

However, if we include back-to-back binding within the theory, we find $\rhotree =
\rhoL - \rhoKK$, where $\rhoKK$ is the number density of back-to-back
bonds.  Both $\rhoL$ and $\rhoKK$ are increasing functions of
$\rho$ 
-- in this case the theory predicts that phase transitions can occur 
only when $\rhotree<0$.  Note however that if
$\rhotree$ is negative then it no longer represents a good estimate of
the number of clusters in the system: due to the construction of
$\rhotree$, this signifies that a proliferation of loops is necessary
for condensation into a liquid phase.

The theory used here treats back-to-back binding
in a schematic way and is therefore not quantitatively accurate.
Nevertheless, this analysis illustrates the importance of cluster
loops in liquid-vapour phase transitions.  By contrast, the phase
transitions that occur in patchy colloids are driven by branching and do not require ``non-specific''
interactions such as the back-to-back binding considered here --
this is possible because there is no difference between the ``inward''
and ``outward'' bonds shown in Fig.~\ref{fig:network}, so there is no constraint on the number
of loops within a single cluster.
In those cases, the Wertheim theory is quantitatively
accurate~\cite{Bianchi2006,Bianchi2008}.

These results for indented colloids illustrate the subtle role of the
bonding topology in hierarchical fluids.  We see that loops are
essential for condensation -- they may arise either from ``undirected''
branching (as in patchy colloids) or from non-specific interactions
(as in polymer-solvent criticality and indented colloids).  Yet
another point of reference is provided by dipolar
fluids~\cite{Safran2000,Camp2007,Rovigatti2013}, where formation of
ring (or loop) structures acts to \emph{suppress} condensation, due to
the absence of non-specific interactions between closed rings, and the
low probability of branching.  Thus, while the formation of colloidal
polymers may appear similar for several kinds of particle, the energy
scales associated with liquid-vapor phase separation can be quite
different, as can the properties of the coexisting phases.

In conclusion, indented colloids with depletion interactions support
unusual liquid and vapor phases, which arise from a hierarchy of
energy scales for lock-key and back-to-back binding.  When the energy
scales are well-separated, these systems also share some properties
with empty liquids~\cite{Bianchi2006,Bianchi2008,Ruzicka:2011ud},
dipolar fluids~\cite{Safran2000,Camp2007,Rovigatti2013}, and
polymer-solvent
systems~\cite{Grosberg,Wilding1996c,Frauenkron:1997ng,pana1998}, but we
emphasise that these systems have different mechanisms of
condensation, due to the different topological properties of the bond
network in each case. For small $q$, the liquid phases of indented colloids are ``porous'',
being characterised by large percolating voids, the typical size of which can
be tuned by varying the properties of the depletant fluid. These states resemble ``equilibrium gels'', which are not
dynamically arrested~\cite{Bianchi2006,Ruzicka:2011ud}.  The fine
experimental control that can be achieved by manipulation of particle shape and depletant properties~\cite{Lekkerkerker:2011}
makes indented colloids ideal systems for further study in this direction.

We are grateful to the EPRSC for support for DJA and NBW through grant
EP/I036192/1 and support for RLJ through grant EP/I003797/1.

\eject

\begin{appendix}

\newcommand{\ee}{\mathrm{e}}

\newcommand{\figpot}{1} 
\newcommand{\figpd}{2} 
\newcommand{\fignet}{3} 

\newcommand{\citeClem}{{[38]}}
\newcommand{\citeAshton}{{[17]}}
\newcommand{\citeWert}{{[34]}}
\newcommand{\citeBianchi}{{[23]}}
\newcommand{\citeWilding}{{[29]}}
\newcommand{\citeBruce}{{[29,30]}}
\newcommand{\citeKim}{{40]}}
\newcommand{\citeHM}{{[39]}}

\section{PARAMETERISATION OF EFFECTIVE POTENTIAL}
\label{sec:effpot}

\subsection{Theoretical motivation}

As described in the main text, we use an effective potential that captures the important features of the
depletion interaction between lock-and-key colloids.  
The effective interaction potential is $v(\bm{r}_{12},\bm{n}_1,\bm{n}_2)$ 
where the vector between two particles is $\bm{r}_{12}$ and their orientations are $\bm{n}_{1,2}$.
Our goal is to parameterise the function $v$
so that our simplified model accurately represents the behaviour of a mixture
of hard indented colloids and ideal depletant particles.  The optimal effective model would match
two-body interactions exactly:
\begin{equation}
\exp[-\beta v(\bm{r}_{12},\bm{n}_1,\bm{n}_2)]  = {\tilde g}_0(\bm{r}_{12},\bm{n}_1,\bm{n}_2)
\label{equ:eff}
\end{equation}
where ${\tilde g}_0$ is the two-particle distribution function for the colloids, normalised so that $\tilde{g}_0=1$ at
large distances.
The subscript `0' indicates that this function is calculated in a dilute limit, where
the colloid density is taken to zero (at fixed depletant volume fraction $\etas$). 

The function ${\tilde g}_0$ can be obtained computationally, but it
depends on the distance $r_{12} = |\bm{r}_{12}|$ and on three different angles.  Parameterising 
this non-trivial function of a four variables is rather difficult -- we have used numerical simulation to identify the most
important features of this function, which we incorporate into a simplified effective potential.

\begin{figure}
\includegraphics[width= 0.75 \columnwidth]{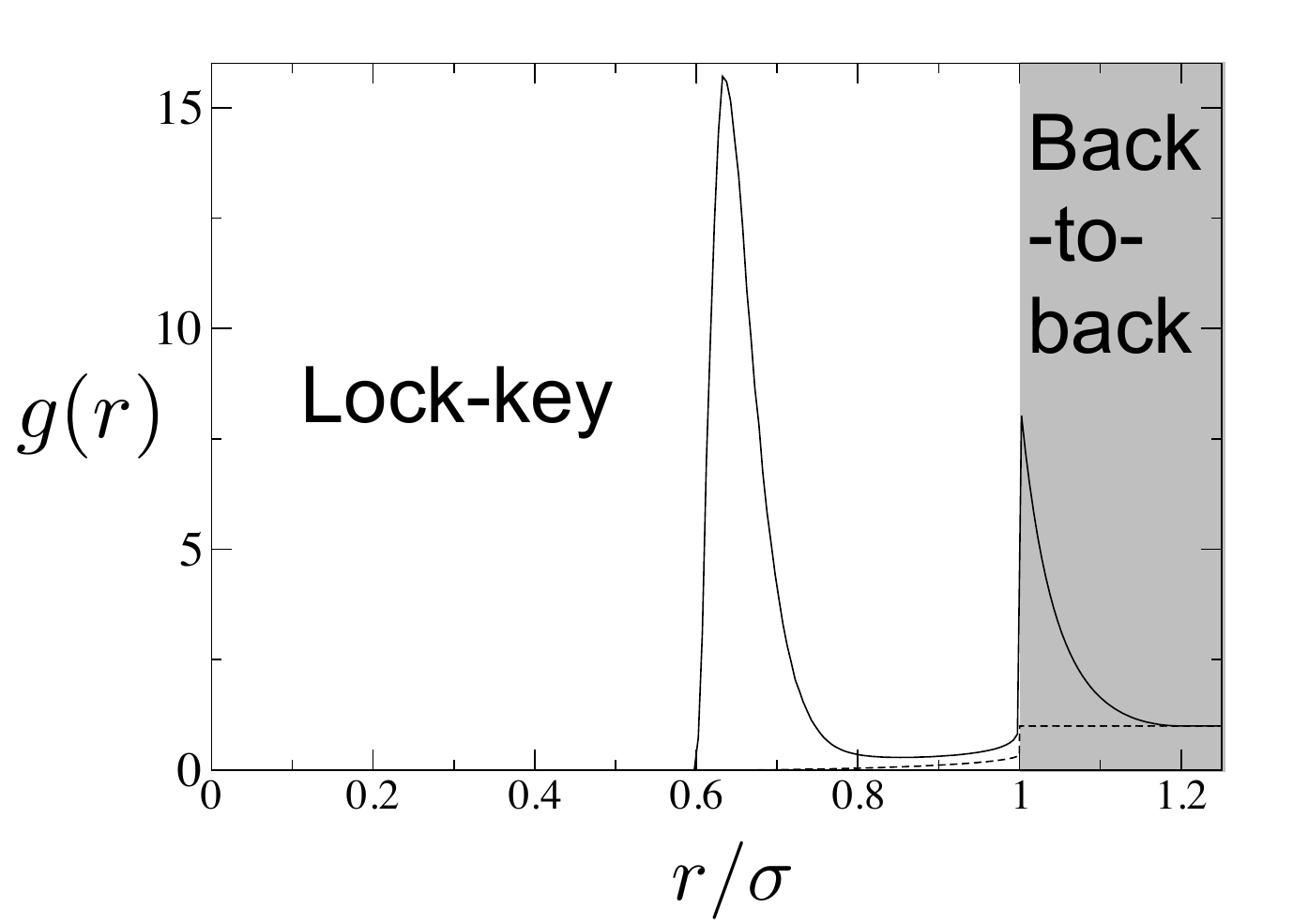}
\caption{Radial distribution function for two indented colloids, with depletant size ratio, $q=0.2$, and $\etas=0.30$.  (Recall $d_c=0.6\sigma$.)
A thin dashed line shows the corresponding function in the absence of depletant. }
\label{fig:grmap}
\end{figure}

The radial distribution function in the dilute limit may be obtained as
\begin{equation}
g_0(r) = \int \frac{\mathrm{d}\bm{n}_1 \mathrm{d}\bm{n}_2}{(4\pi)^2}\, {\tilde g}_0(\bm{r}_{12},\bm{n}_1,\bm{n}_2)
\label{equ:gr}
\end{equation}
where both integrals run over the entire unit sphere.  Fig.~\ref{fig:grmap} shows an example of this function,
in a situation where both lock-key and back-to-back binding are signficant.  The two separate peaks indicate
the different binding mechanisms, but this representation does not reveal the orientation dependence of the
interaction.

\begin{figure}
\begin{center}
\includegraphics[width= \columnwidth]{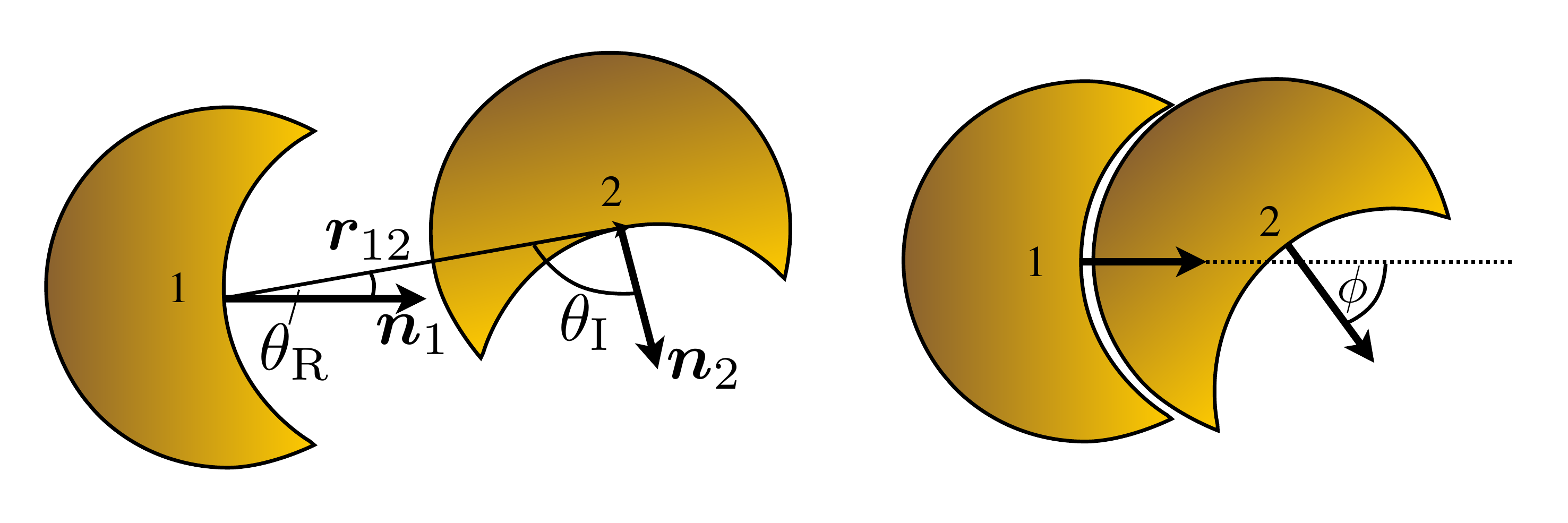}
\caption{Co-ordinate system for interactions among indented colloids. The angles between the interparticle vector $\bm{r}_{12}$ and the directors $\bm{n}_{1,2}$
are labelled as shown: the labels $\theta_{\rm R,I}$ are chosen such that $\theta_{\rm R} < \theta_{\rm I}$.
The angle between the directors, is $\phi$.}
\label{fig:coords}
\end{center}
\end{figure}

Fig.~\ref{fig:coords} shows the choice of co-ordinate system that we have used to capture this dependence.  In lock-key binding, one particle
plays the role of the ``lock'' and the other the ``key''.  The definition $\cos\theta_{\rm R} = 
\max( \bm{n}_1\cdot \bm{r}_{12},-\bm{n}_2\cdot \bm{r}_{12})$ takes the smaller of the angles
between $\bm{r}_{12}$ and the directors $\bm{n}_{1,2}$; the particle associated with this smaller angle plays the part of
the lock.  We define $\cos\theta_{\rm R}^c = d_c/\sigma$ so that $\theta_{\rm R}^c$ corresponds to angle
between the director and a vector pointing out through the ``lip'' of the lock mouth.  Lock-and-key binding happens only for $\theta_{\rm R}<\theta_{\rm R}^c$
while back-to-back binding happens for $\theta_{\rm R}>\theta_{\rm R}^c$.  

Since back-to-back binding takes place between two convex surfaces, this part of the interaction depends weakly on angular
co-ordinates, so the effective potential includes a square well of range $r_{\rm BB} = 1.1\sigma$ in that case.  The contribution
to the effective potential $v$ is therefore $v_{\rm BB} =-\varepsilon_{\rm BB} \Theta( \theta_{\rm R} - \theta_{\rm R}^c) \Theta(1.1\sigma-r_{12})$
where $\Theta(x)$ is the step function.
 For lock-and-key
bonding, the angular dependence is less simple, but the dominant effect can be accounted for through the angle $\phi$,
which measures the rotation of the ``key'' particle within the mouth of the lock.  Hence the interaction
potential for lock-key binding is $v_{\rm LK}=-\varepsilon_{\rm LK} W(\phi) \Theta( \theta_{\rm R}^c - \theta_{\rm R}) \Theta(d_c+0.1\sigma-r_{12})$ where  $W(\phi)$ is the piecewise linear function of $\cos\phi$ shown in Fig.~\figpot~of the main text.
In fact, this simple function provides an almost quantitative description of the $\phi$-dependence of the effective potential~\citeClem.

\subsection{Numerical parameterisation}

Having specified the form of the effective potential, we now describe how the parameters $(\epsLK,\epsBB)$ are chosen.
The (effective) second virial coefficient for the indented colloids in the presence of the depletant is~\citeHM
\begin{equation}
B_2 = 2\pi
\int_0^\infty \mathrm{d}r\, r^2 [1-g_0(r)]
\end{equation}
We split $B_2$ into two pieces, associated with lock-key and back-to-back binding:
\begin{align}
B_2^{\rm LK} &= 2\pi
\int_0^{\sigma} \mathrm{d}r\, r^2 [1-g_0(r)]
\nonumber \\
B_2^{\rm BB} &= 2\pi
\int_{\sigma}^\infty \mathrm{d}r\, r^2 [1-g_0(r)]
\label{equ:B2split}
\end{align}
These two parameters indicate the strength of the two kinds of binding.  A representative plot of $g_0(r)$
is shown in Fig.~\ref{fig:grmap}.  The peak at $r\approx d_c$ is associated with lock-key binding, and its area sets
the strength of this interaction.  The peak at $r\approx\sigma$ is associated with back-to-back binding.

From (\ref{equ:gr}), $B_2^{\rm LK}$ and $B_2^{\rm BB}$ can be written as integrals of $1-\tilde{g}_0$.  Motivated by (\ref{equ:eff}), 
our procedure is therefore to choose $\epsLK,\epsBB$ so that the corresponding integrals of 
$1-\exp[-\beta v(\bm{r}_{12},\bm{n}_1,\bm{n}_2)]$ are equal to $B_2^{\rm LK},B_2^{\rm BB}$.  These integrals
are contributions to the second virial coefficient of the effective model, so we denote them
by $B_{2,{\rm eff}}^{\rm LK}$ and $B_{2,{\rm eff}}^{\rm BB}$.

Since the potential is a square well and the hard-particle interactions have range at most $\sigma$, it is easily verified
that
\begin{align}
B_{2,{\rm eff}}^{\rm LK} &= v_{\rm m} - v_{\rm LK} ( 1 - \ee^{\epsLK} )
\nonumber \\
B_{2,{\rm eff}}^{\rm BB} &=  v_{\rm BB} ( 1 - \ee^{\epsBB} )
\label{equ:b2eff}
\end{align}
where $v_{\rm m}$ accounts for the excluded volume between two indented colloids, 
and $v_{\rm LK},v_{\rm BB}$ are phase space volumes associated with
the two binding modes.

\begin{figure}
\includegraphics[width= 0.75 \columnwidth]{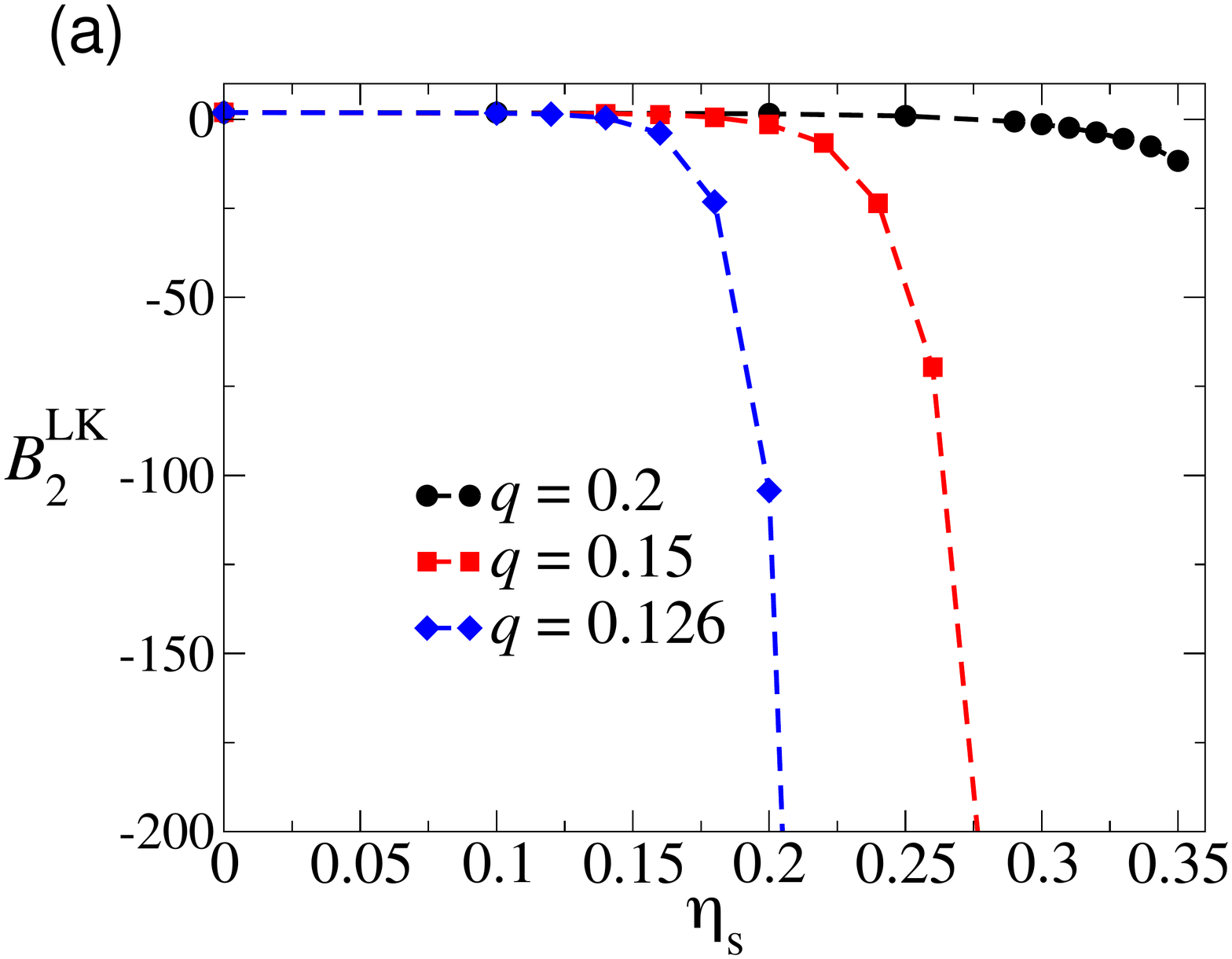}
\includegraphics[width= 0.75 \columnwidth]{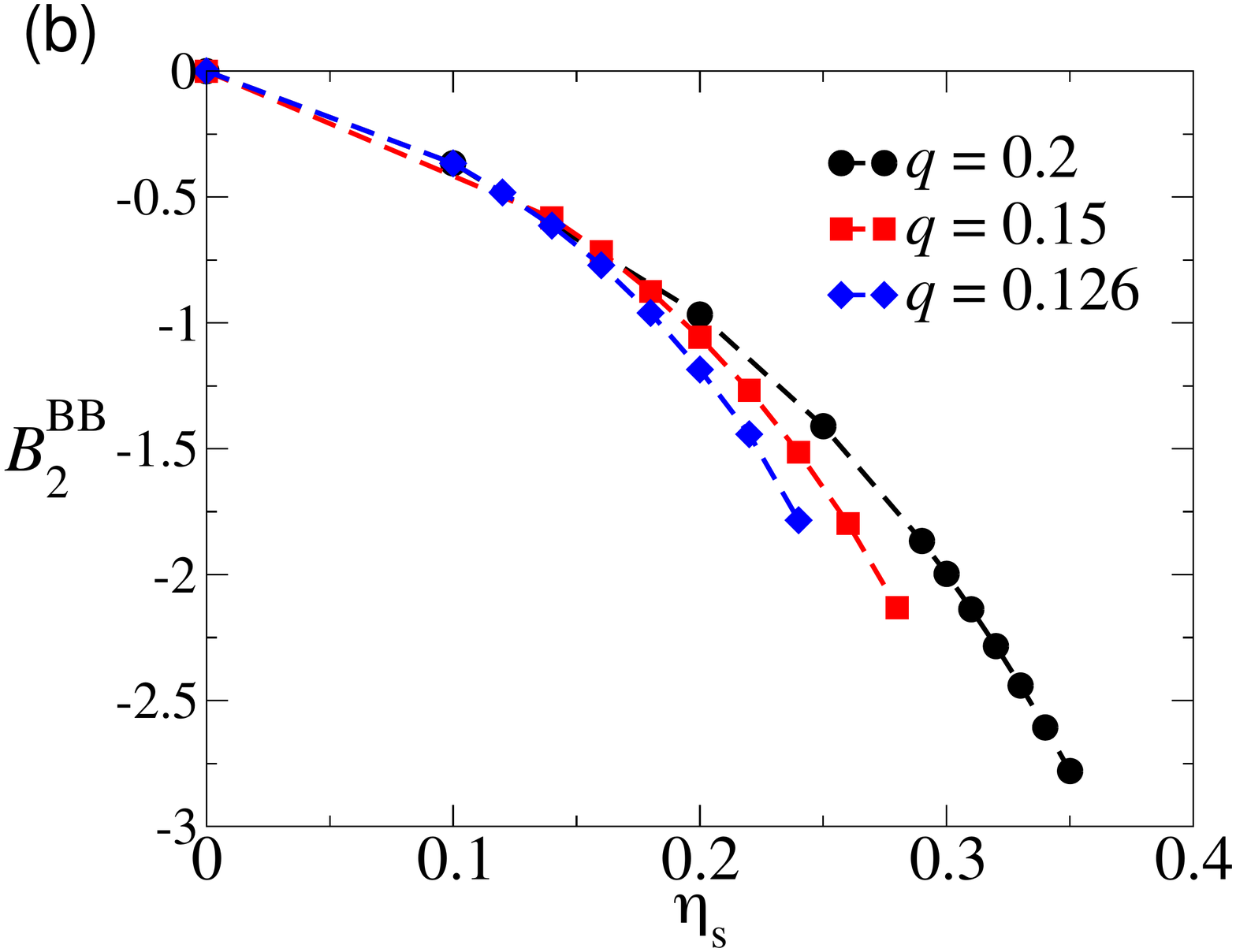}
\caption{The contributions to the second virial coefficient for a system of two locks, interacting with a depletant of penetrable spheres.}
\label{fig:etalb2}
\end{figure}

\begin{figure}
\includegraphics[width= 0.75 \columnwidth]{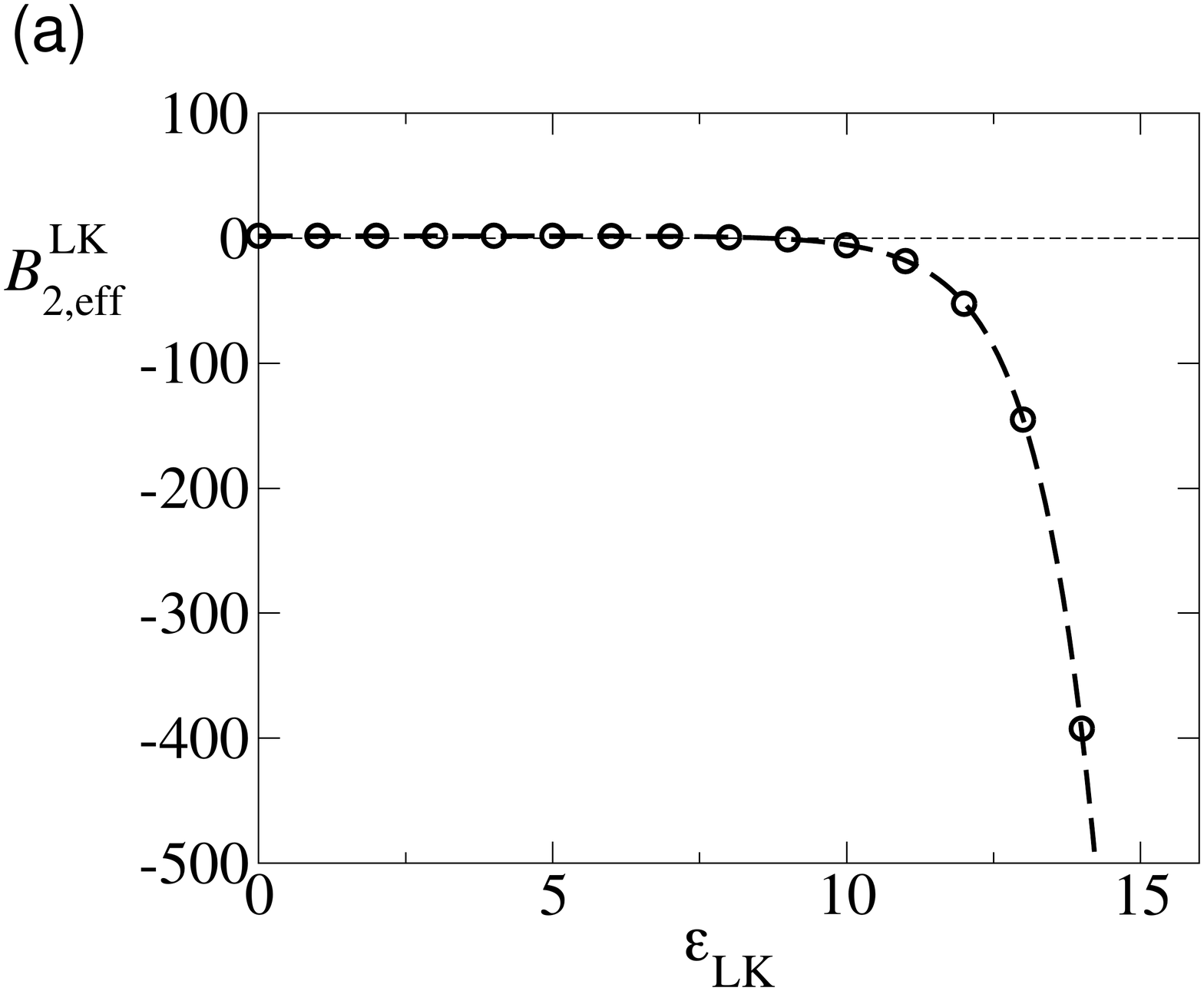}
\includegraphics[width= 0.75 \columnwidth]{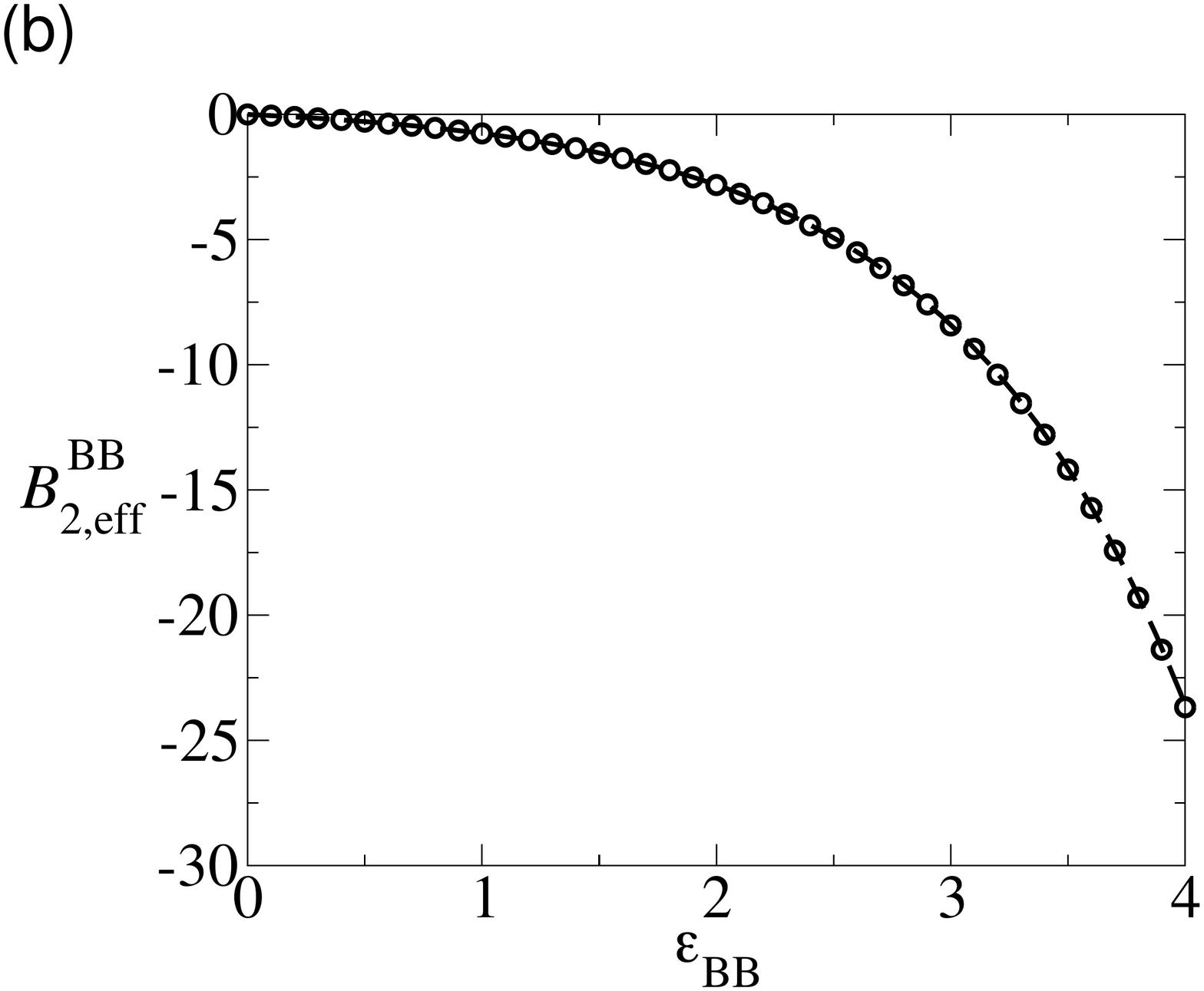}
\caption{Contributions [see (\ref{equ:B2split})] to the second virial coefficient for a system of two indented colloids interacting via the effective potential.  
Dashed lines are fits to (\ref{equ:b2eff}) with $v_{\rm m}=1.94\sigma^3$, $v_{\rm LK}=(3.3\times10^{-4})\sigma^3$ and $v_{\rm BB}=0.44\sigma^3$}
\label{fig:lb2}
\end{figure}

\subsection{Computational implementation}

To estimate $g_0(r)$ for the system with depletant, we use the Geometrical Cluster Algorithm (GCA) to simulate two colloidal
particles in a box, mixed with depletant, as in~\citeAshton~(this procedure accounts for finite-size corrections to $g(r)$).
Representative results for $B_2^{\rm LK},B_2^{\rm BB}$ are shown in Fig.~\ref{fig:etalb2}, for several size ratios $q$.
As expected, these parameters become large and negative as the nanoparticle volume fraction $\etas$ increases.

The corresponding quantities $B_{2,{\rm eff}}^{\rm LK},B_{2,{\rm eff}}^{\rm BB}$ were calculated numerically
by the same method, but using the effective potential instead of an explicit depletant.  Results are shown in Fig.~\ref{fig:lb2}, including
fits to (\ref{equ:b2eff}), from which we estimate the small volume associated with lock-and-key binding to be $v_{\rm LK} = (3.3\times 10^{-4})\sigma^3$.  
Comparison of Figs~\ref{fig:etalb2} and~\ref{fig:lb2} allows straightforward mapping from the depletion
parameters $(q,\etas)$ to the parameters $(\epsLK,\epsBB)$ of the effective potential.

\section{SIMULATION METHOD}
\label{sec:sim} 
\subsection{Biased insertions in the grand canonical ensemble}

The lock-key interactions in our system are typically strong, implying
a strongly negative chemical potential and consequently a very low
acceptance rate for standard particle transfers (insertions and
deletions) in grand canonical Monte Carlo (GCMC) simulations. To counter this we introduce an additional biased transfer
update in addition to the standard one. This operates by 
preferentially attempting to insert an indented particles in a bound configuration around a target colloid.

To implement the ``biased insertion'', we first pick a target colloid randomly from
the existing ones in the box. The location of the centre of the colloid to be
inserted is then chosen randomly from within a narrow spherical shell
surrounding the target. The shell radii are chosen to represent the
typical separation of bound colloids. For a given point in the shell
there will be a range of solid angle for the particle director which
avoids hard particle overlaps, and we choose the director of the new
particle uniformly from within this range. The biasing factor needed
to maintain detailed balance is therefore a product of the shell
volume and the solid angle that avoids overlap. This biasing factor
can also be calculated for a bound particle which is to be removed
from the system, based on its separation from its bound neighbour, and
so enters symmetrically in the grand canonical acceptance
probabilities.

\subsection{Determination of binodals, critical points and field mixing}

\begin{figure}
\begin{center}
\includegraphics[width= 0.98 \columnwidth]{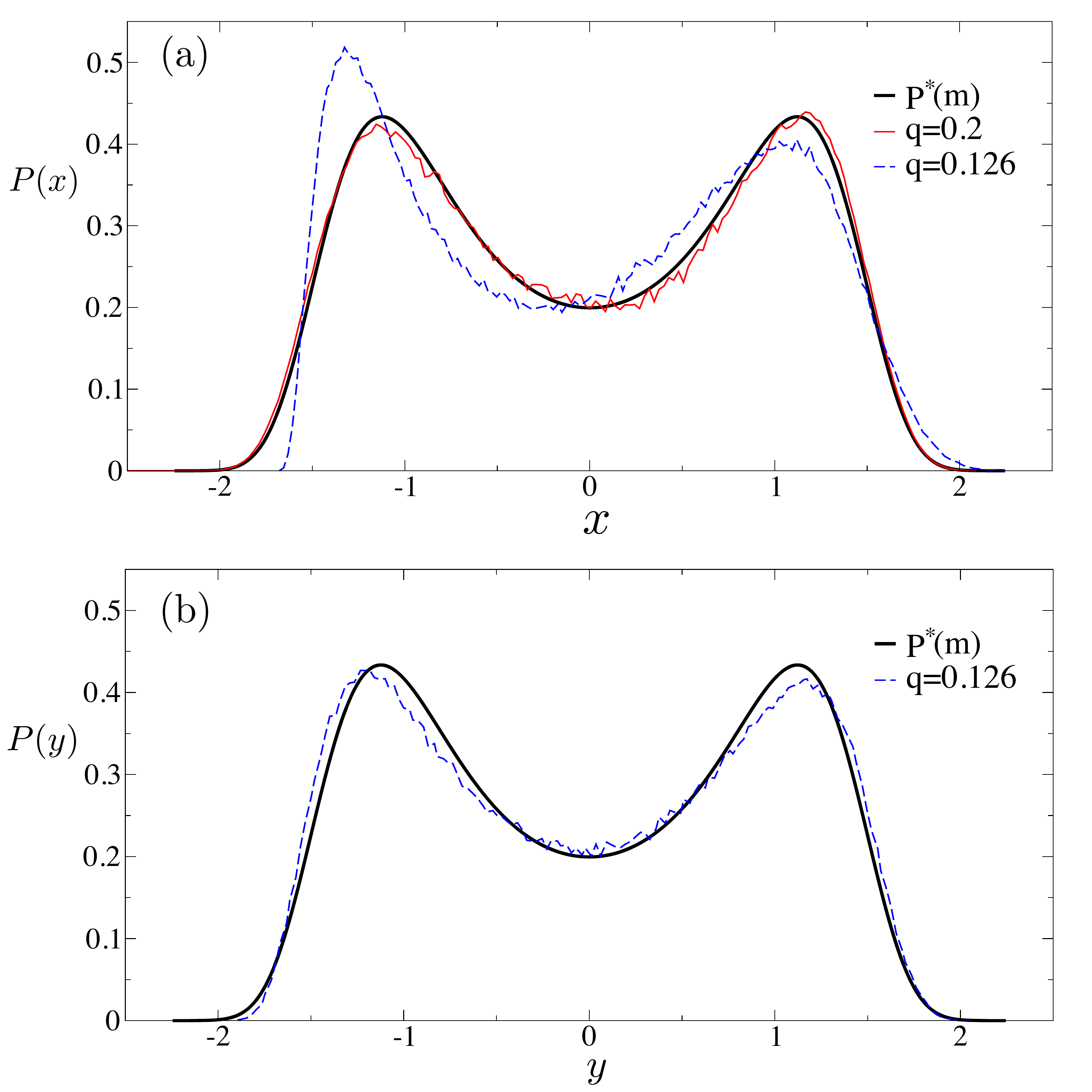}

\caption{(a) Critical point order parameter distributions $P(x)$ with
  $x=a(\rho-\rho_c))$. The figure shows data for $q=0.2$ and $q=0.126$
  and the distributions are scaled to unit norm and variance via the
  choice of the parameter $a$. Also shown for comparison is the fixed
  point Ising magnetisation distribution $P^\star(m)$.  (b) The
  corresponding normalised distribution of the field mixed order
  parameter $P(y)$ with $y=b({\cal M}-{\cal M}_c)$ and $b$ chosen to
  scale the distributions to unit variance. Box sizes were
  $\ell=12\sigma$ for the $q=0.126$ system and $\ell=7.5\sigma$ for
  the $q=0.2$ system.}

\label{fig:critical}
\end{center}
\end{figure}

Our GCMC simulations sample the fluctuations in the colloid number density
$\rho=N/V$ and energy density $u=E/V$. Well within the two-phase
region, the probability distribution $P(\rho)$ exhibits a pair of
widely separated peaks centered on the coexistence
densities. Accordingly $\rho$ serves as an order parameter for the
transition, and the coexistence densities (binodal) can be simply read
off from the peak positions~\citeWilding. Closer to the
liquid-vapor critical point, the fluctuations in $\rho$ and $u$ assume
a universal character which can be probed via the properties of the
joint distribution $P(\rho,u)$~\citeBruce. In models
that exhibit particle hole symmetry, the order parameter distribution
at criticality matches the (independently known) universal fixed point
form of the Ising magnetisation distribution $P^\star(m)$. Effecting
this matching provides a means of estimating the critical
parameters. However, since our model lacks particle-hole symmetry,
field mixing is expected to occur. This implies that the scaling
equivalent of $m$ is the linear combination ${\cal M}=\rho-su$
\citeBruce. (Strictly speaking, a non-linear combination
may be appropriate~\citeKim, but the linear combination 
is a simple and practical choice.).  To estimate the critical
parameters the field mixing parameter $s$ must be tuned along with
$\mu$ and $T$ such as to best match $P({\cal M})$ to
$P^\star(\tilde{m})$ \citeWilding.

We have used these techniques to determine the critical points and
coexistence binodals for a selection of values of $q$ in the interval
$q=[0.25,0.1]$; the results are shown in Fig.~\figpd(a,d)~of the main
text. Our universal distribution matching method also delivers
estimates for the field mixing properties of the critical fluid. For
$q=0.2$, a satisfactory matching of the number density distribution to
the Ising fixed point form can be obtained as shown in Fig.~\ref{fig:critical}(a),
so field mixing is negligible in this case. By contrast, for
$q=0.126$, no satisfactory matching can be obtained under the assumption $s=0$.
The situation is greatly improved, however,  by forming the distribution
of the mixed field ordering operator $P({\cal M})$, as shown in
Fig.~\ref{fig:critical}(b), which corresponds to  
$s=-0.058(1)$

\section{VOID SIZES IN LIQUID STATES}
\label{sec:void}

\begin{figure}
\includegraphics[width=0.99\columnwidth]{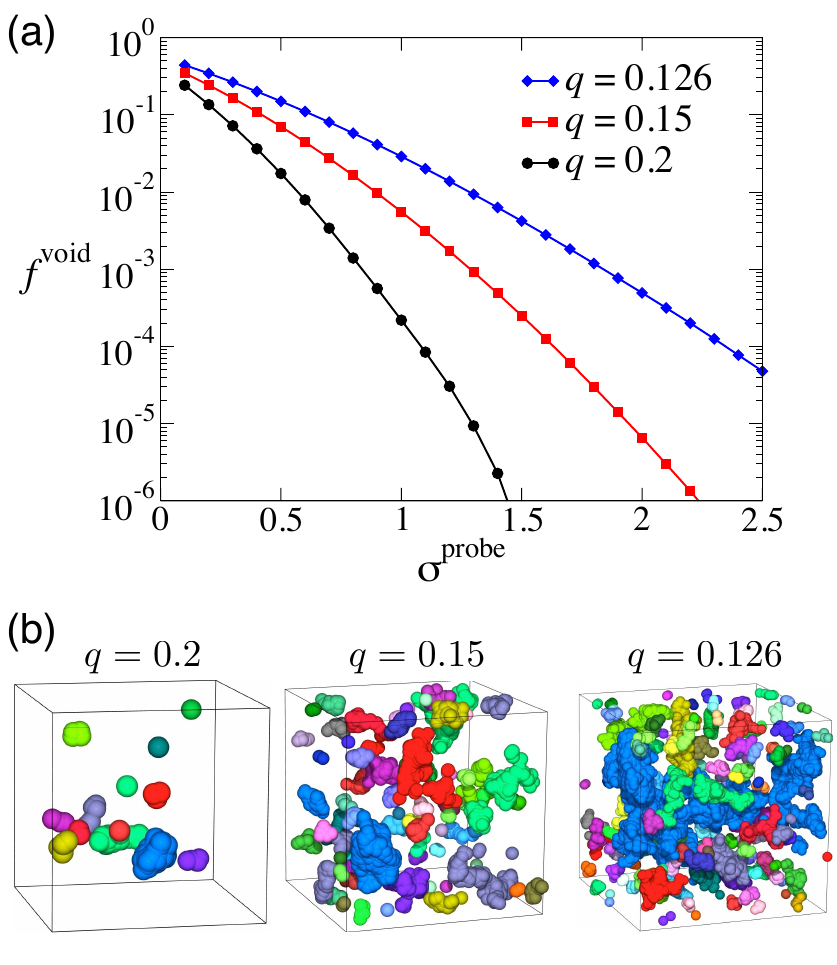}
\caption{(a) Fraction of space $f^{\rm void}$  available to spherical ``ghost particles'' of size $\sigma^{\rm probe}$, in liquid states
at different size ratios $q$. (The probe diameter is measured in units of the colloid diameter $\sigma$.) For $\sigma^{\rm probe}=1$,
the porous liquids at $q=0.126$ have many more cavities where ghost particles will fit, compared with the dense liquids at $q=0.2$.
(b)~Depiction of void spaces in representative configurations, for $\sigma^{\rm probe}=1$.  Each connected void space is shown with a different
color, to illustrate their sizes and connectivity.}
\label{fig:void}
\end{figure}

In Fig.~\ref{fig:void}, we show the fraction $f^{\rm void}$ of the total volume of the system that is accessible to ``probe particles'' (or ``ghost'' particles) of size
$\sigma^{\rm probe}$.  These calculations were performed at states where $\etas=1.05\etas^c$. 
The porous structure of the liquid at $q=0.126$ means that
$f^{\rm void}$ is greater by several orders of magnitude than the corresponding state at $q=0.2$.  We also show the void spaces
associated with typical configuration, at $\sigma^{\rm probe}=1$.  Note that the sizes of the typical voids increase along with their
volume fraction; at $q=0.126$, the largest cluster has percolated the system.

\section{WERTHEIM THEORY}
\label{sec:wert}

In the main text, arguments based on Wertheim's theory of associating fluids~\citeWert~were used to investigate the effects
of branching and non-specific binding on phase behaviour.
The theory consists of two main parts.  The first is an exact representation
of a fluid's partition function, where the density is decomposed as a sum of partial densities that 
describe particles in different local environments.  The second part is an (approximate) thermodynamic perturbation
theory (TPT) which facilitates explicit calculations.
In \citeAshton, we developed this theory for a system of indented colloids that forms linear chains, but both chain branching
and back-to-back binding were neglected.  

The generalisation of this theory to include these effects is straightforward, although
some book-keeping is required to keep track of the various local environments.  Briefly, the interaction between particles
is broken down into three parts, which take care of particle overlaps, lock-and-key binding, and back-to-back interactions.  This decomposition
is performed at the level of the Mayer $f$-function, so the diagrammatic expansion of the free energy contains three kinds of interaction.

Within this diagrammatic expansion, Wertheim's theory requires that particles be classified according to their local bonding.  In practice, 
we keep track of the number of inward and and outward lock-and-key bonds, and the number of back-to-back bonds.  The constraint (recall Fig.~\fignet~of main text) that
each particle has at most one outward bond is enforced at the level of the diagrammatic series (this is the key advantage of Wertheim's approach).

%

\subsection{Quasispecies densities and mass-action equations} 

To develop the theory, let $\rho_{ijk}$ be the number of particle with $i$ inward bonds, $j$ outward bonds and $k$ back-to-back bonds.  These are the quasispecies
densities of Wertheim's theory~\citeWert.  For simplicity we assume that particles may have at most two inward
lock-and-key bonds ($k\leq2$), at most one outward lock-and-key bond ($i\leq 1$) and at most one back-to-back bond ($k\leq 1$).
Further, we assume that
that particles with back-to-back bonds have no inward lock-and-key bonds and vice versa, so that
only one of $j$ or $k$ may be greater than zero.
Even with these constraints, there are still eight distinct local environments, but
the advantage of the TPT is that
the various $\rho_{ijk}$ have fairly simple relations between them.

We postpone the detailed analysis to a later work and quote only the main results, with their physical interpretation.  We
use the notation $\rho_0=\rho_{000}$, for brevity.  ``Chemical equilibrium'' between
free monomers and particles with a single inward bond (and no other bonds) means that
\begin{equation}
\rho_{010} = K \rho_{0} \rhoL
\label{equ:rho010}
\end{equation}
where
\begin{equation}
\rhoL = \rho_0 + \rho_{100} + \rho_{020} + \rho_{001}
\end{equation}
is the density of unoccupied lock sites (see also the main text),
and $K$ is an equilbrium constant which, at the simplest level, is equal to $B_{2,\rm eff}^{\rm LK}$.  More generally, $K$
 may be obtained within the TPT as an integral involving the radial distribution function $g(r)$ in a hard-particle
reference system.

We also define
\begin{align}
\rhoK & = \rho_0 + \rho_{010} \nonumber \\
\rhoLK & = \rho_{010} + \rho_{110} \nonumber  \\
\rhoLLK & = \rho_{020} + \rho_{120} 
\end{align}
which correspond to number densities of key surfaces with $0$, $1$, and $2$ inward bonds.  Also,
\begin{equation}
\rhoKK  = \tfrac12 ( \rho_{001} + \rho_{101})
\end{equation}
is the number density of back-to-back bonds.

Chemical equilibria between other particle quasispecies yield
\begin{align}
\rho_{001} &= K_{\rm BB} \rho_{0} \rhoK \nonumber  \\
\rho_{020} &= a K^2 \rho_{0} \rhoL^2 
\label{equ:rho001}
\end{align}
where $K_{\rm BB}$ is an equilibrium constant for back-to-back binding, and $a$ is a geometrical factor that accounts
for the reduced volume available when binding a second lock onto a key where a lock is already bound.  (The theory
provides integral expressions for these quantities; in general we expect $a$ to depend weakly on
the energy scales $\epsLK,\epsBB$ while $K_{\rm BB}\propto \ee^{\beta\epsBB}$.)
The Wertheim theory also builds in independence of bonds for the lock- and key-sites of a given 
particle.  This means for example that $\rho_{110} = \rho_{100}\rho_{010}/\rho_0$, and similarly
$\rho_{120} = \rho_{100}\rho_{020}/\rho_0$ and $\rho_{101} = \rho_{100}\rho_{001}/\rho_0$.  

Combining all the definitions and theoretical results given so far, we arrive at
\begin{align}
\rhoLK & = K \rhoL \rhoK \nonumber \\
\rhoLLK & = a K \rhoL^2 \rhoK \nonumber \\
\rhoKK & = K_{\rm BB} \rhoK^2 
\label{equ:ma}
\end{align}
which are the ``mass-action'' equations described in the main text.

Since the total density $\rho$ is equal to the sum of all the partial densities $\rho_0,\rho_{001}$ etc, we also have
\begin{align}
\rho & =  \rhoK + \rhoLK + \rhoLLK + 2\rhoKK
\label{equ:key-sum}
\end{align}
which simply means that all key-sites must be either free or involved in some kind of bonding. Combining (\ref{equ:rho010},\ref{equ:rho001})
yields $\rho_{100}\rhoL = (\rho_{010} + 2\rho_{020})\rhoK$ from which one may derive
\begin{equation}
\rho = \rhoL + \rhoLK + 2\rhoLLK
\label{equ:lock-sum}
\end{equation}
This equation enforces the constraint that all lock-sites are either occupied or unoccupied, analogous to (\ref{equ:key-sum}).

\subsection{Pressure}

Following~\citeBianchi, we obtain the pressure
of the system within the TPT as a sum of two terms
\begin{equation}
\beta P = \rho_{\rm tree} + \Delta P_{\rm ref}
\end{equation}
where $\rho_{\rm tree}$ is an estimate of the number of clusters in the system and $\Delta P_{\rm ref} = \beta P_{\rm ref} - \rho$ is
a contribution from the hard-particle reference system, which has a reduced pressure $\beta P_{\rm ref}$ that depends on the total
density $\rho$ but not on the quasispecies densities.  As discussed in the main text, $\rho_{\rm tree}$ is equal to the number of
clusters in the system if there are no internal loops within clusters (in that case, formation of any bond decreases the total number of clusters by $1$).
This ``tree-like'' assumption motivates the notation $\rho_{\rm tree}$.

The theory described here gives 
\begin{equation}
\rho_{\rm tree} = \rhoL - \rhoKK .
\label{equ:tree}
\end{equation}
To see this, note that if all clusters are tree-like then 
their number is equal to the difference between the total number of particles and the total number of bonds (either lock-key
or back-to-back).  The number of lock-key bonds is $\rhoLK + 2\rhoLKK = \rho-\rhoL$ and the number of back-to-back bonds
is $\rhoKK$, yielding (\ref{equ:tree}).

\subsubsection{Conditions for phase separation}
\label{sec:sep}

Liquid vapour phase transitions occur if $(\partial P/\partial V)_{N,T}>0$ for some volume $V$.  Here the derivative is taken
at a constant value of the total particle number $N$, but the numbers of particles within each quasispecies depend on $V$.
The reference contribution $\Delta P_{\rm ref}$ is decreasing in $V$ so a necessary condition for phase separation within Wertheim's theory is
that $(\partial \rho_{\rm tree}/\partial V)_{N,T}>0$.  At constant $T$, $\rho_{\rm tree}$ is a simple function of $\rho$, which implies
$V(\partial \rho_{\rm tree}/\partial V)_{N,T} = -\rho(\mathrm{d}\rho_{\rm tree}/\mathrm{d}\rho)$, so phase separation can
occur only if 
\begin{equation}
(\mathrm{d}\rho_{\rm tree}/\mathrm{d}\rho) < 0.
\label{equ:sep-suff}
\end{equation}
(More precisely, one requires $\mathrm{d}\rho_{\rm tree}/\mathrm{d}\rho < 
-\mathrm{d}(\Delta P_{\rm ref})/\mathrm{d}\rho$.)

The relation (\ref{equ:sep-suff}) provides a useful necessary condition for phase separation. 
In particular, for the system considered here, the densities $\rhoL$, $\rhoK$, $\rhoKK$, etc are
all monotonic in $\rho$ (see Sec.~\ref{sec:mono} below).  In the absence of back-to-back binding,
one has $\rhotree=\rhoL$ from (\ref{equ:tree}), the pressure is monotonic in $\rho$, so
the theory predicts that no phase transition is possible.  

Physically, the idea is that internal loops within bonded clusters of particles
are suppressed by the directed lock-key bond and the fact that a lock-site can bind to
 at most one key-site.   The Wertheim theory captures this property.
 
 \subsubsection{Monotonicity of $\rhoL$, $\rhoK$}
\label{sec:mono}

To complete the argument that phase transitions within the theory requires $K_{\rm BB}>0$, we prove
that $\rhoL$ and $\rhoK$ depend monotonically on $\rho$.  
Starting from (\ref{equ:lock-sum}), we use (\ref{equ:ma}) to express all quantities in terms of $\rhoL,\rhoK$, and
we consider small variations in $(\rho,\rhoL,\rhoK)$, obtaining
\begin{equation}
\delta \rho = \delta\rhoL ( 1 + K\rhoK + 4aK^2\rhoL\rhoK) + \delta\rhoK ( K\rhoL + 2 a K^2 \rhoL^2 )
\label{equ:dr}
\end{equation}
 From (\ref{equ:key-sum},\ref{equ:lock-sum}), we have
\begin{equation}
\rhoL + \rhoLLK = \rhoK + 2 \rhoKK
\label{equ:sum-both}
\end{equation}
and again considering small variations in $\rhoL,\rhoK$ yields
\begin{equation}
\delta\rhoL ( 1 + 2 aK^2 \rhoL\rhoK) = \delta \rhoK ( 1 - a K^2 \rhoL^2 + 4 K_{\rm BB} \rhoK ).
\end{equation}
Using (\ref{equ:sum-both}) again gives
\begin{equation}
\delta\rhoL ( 1 + 2 aK^2 \rhoL\rhoK) = \frac{\delta \rhoK}{\rhoK} ( K\rhoL + 2K_{\rm BB}\rhoK ).
\label{equ:drlk}
\end{equation}
It follows from (\ref{equ:dr},\ref{equ:drlk}) that $\delta\rho$ and $\delta\rhoK$ are both monotonic in $\delta\rhoL$,
which implies that $(\rhoL,\rhoK)$ are both monotonic in $\rho$, as are $(\rhoKK,\rhoLK,\rhoLKK)$.  Given that
$\rho_{\rm tree} = \rhoL - \rhoKK$, it follows that $\rhotree$ must be monotonic if $\rhoKK=0$, in 
which case $(\mathrm{d}\rho_{\rm tree}/\mathrm{d}\rho) > 0$ and the theory predicts no phase transition.

\end{appendix}

\end{document}